%% file: twin_spine_rx.tex
\documentclass[11pt,a4paper]{article}
\usepackage[T1]{fontenc}
\usepackage{amsmath,amssymb,geometry,xcolor,booktabs,longtable,array,graphicx}
\usepackage{listings}
\usepackage[most]{tcolorbox}
\usepackage[colorlinks=true,linkcolor=blue!60!black,citecolor=blue!60!black,urlcolor=blue!60!black]{hyperref}
\input{build/values.tex}

\makeatletter
\def\stcol@IDENTITY{teal!70!black}
\def\stcol@POSTULATE{orange!80!black}
\def\stcol@IDENTIFIED{violet!80!black}
\def\stcol@DERIVED{green!45!black}
\def\stcol@OPEN{red!85!black}
\DeclareRobustCommand{\cl}[1]{{\edef\@st{\csname clst@#1\endcsname}%
  \textcolor{\csname stcol@\@st\endcsname}{\textsf{\footnotesize[#1\,\@st]}}}}
\newcommand{\st}[1]{\textcolor{\csname stcol@#1\endcsname}{\textsf{\footnotesize #1}}}
\makeatother

\lstdefinestyle{spine}{language=Python,basicstyle=\ttfamily\footnotesize,
  keywordstyle=\color{blue!60!black},commentstyle=\color{gray!70!black}\itshape,
  stringstyle=\color{red!50!black},columns=fullflexible,breaklines=true,
  rangeprefix=\#\ [,rangesuffix=],includerangemarker=false,gobble=4,
  showstringspaces=false,upquote=true}
\newtcolorbox{mathbox}[1][]{enhanced,breakable,colback=white,colframe=gray!40,boxrule=0.5pt,
  left=6pt,right=6pt,top=4pt,bottom=4pt,before skip=2pt,
  title={\footnotesize\textsf{In standard notation}},coltitle=black,colbacktitle=gray!10,
  attach boxed title to top right={yshift=-\tcboxedtitleheight/2,xshift=-4mm},
  boxed title style={boxrule=0.5pt,colframe=gray!40,colback=white},#1}
\newcommand{\spine}[2]{%
  \begin{tcolorbox}[enhanced,breakable,colback=gray!4,colframe=gray!45,boxrule=0.4pt,
    left=4pt,right=4pt,top=2pt,bottom=2pt,
    title={\footnotesize\ttfamily twin\_spine.py\ \ [#1]\hfill\normalfont #2},
    coltitle=black,colbacktitle=gray!15]
  \lstinputlisting[style=spine,linerange=#1-/#1]{twin_spine_rx.py}
  \end{tcolorbox}}

\newcommand{\C}{\mathbf C}
\newcommand{\D}{\mathbf D}
\newcommand{\E}{\mathbf E}
\newcommand{\Hv}{\mathbf H}

\title{\textbf{A Candidate twin-spine photon model in MacCullagh's medium}\\[0.3em]
\large Line defects, birefringent torque without superposition, and emission}
\author{%
	B. Hou\\[0.2em]
	{\small Department of Physics and Astronomy, Cardiff University, Cardiff, United Kingdom}\\
	{\small\texttt{HouB6@cardiff.ac.uk}}
	\and
	D. A. Sinclair\\[0.2em]
	{\small Imense Ltd, Cambridge, United Kingdom}\\
	{\small\texttt{david@s-hull.org}}
}
\date{November 2026}

\begin{document}
\maketitle
{\renewcommand{\thefootnote}{}\footnotetext{\copyright\ 2026 Imense Ltd
(registered in England and Wales, company no.\ 05136978; registered office
115c Milton Road, Cambridge CB4~1XE). The text of this article is distributed by
arXiv under the arXiv.org perpetual, non-exclusive licence~1.0; all other rights
reserved. The software \texttt{twin\_spine\_rx.py} and \texttt{make\_figures.py}, the
code excerpts reproduced from them in this article, and the claim ledger generated by \texttt{twin\_spine\_rx.py}
(Appendix~\ref{app:ledger}) are licensed under the GNU Affero General Public
License, version~3 or later (SPDX: \texttt{AGPL-3.0-or-later}).}}

\begin{abstract}\noindent

  We formulate the vacuum as a rotationally elastic continuum, serving as the low-energy,
  purely electromagnetic approximation to the deformable spacetime metric of General Relativity.
  Adopting MacCullagh's single-parameter action affords remarkable mathematical economy and analytical transparency;
  rather than treating electric and magnetic fields as separate axiomatic entities,
  it yields the source-free Maxwell equations directly from the continuum's resistance to twist.
  Within this geometric framework, we demonstrate that a transversely localized field moving rigidly at $c$ requires line defects, the minimal finite arrangement being a twin-spine photon comprising two opposite displacement-current cores. This structure propagates losslessly with zero transverse stress, naturally resolving into linear and circular geometries carrying angular momentum $0$ or $\pm\hbar$. Rather than relying on the phenomenological superposition of phase-delayed plane waves, photon behavior in a birefringent crystal emerges kinematically: the differential wave speed dynamically reshapes the single shared path of the spines, generating an exact, continuous macroscopic torque on the crystal lattice. Furthermore, we provide a mechanical derivation for emission: by modeling the electron as a bound topological defect subject to a strict temporal phase condition (a $\pi$-twist), its transverse acceleration physically shears its co-moving bow wave. The vacuum wave equation forces this residual dipole strain precisely into the twin-spine geometry, while the electron's topological phase lock provides the exact boundary condition required to quantize the emitted wave. Every mathematical step is verified in an accompanying SymPy Executable Research Record, establishing a strict, machine-readable boundary between postulated topology and derived mechanics.

\end{abstract}

%=====================================================================================
\section{Introduction}
%=====================================================================================

Localized models of the photon have a long history alongside the plane-wave description of optics \cite{brittingham1983,ziolkowski1985,bialynicki1994}.
In optical tweezers, light aligns and spins birefringent particles \cite{beth1936,friese1998}.
The standard account of this phenomenon resolves each photon into two orthogonal polarization components that travel at different phase velocities and phenomenologically recombine. This paper presents a single-object account.
It builds a localized photon within the rotationally elastic continuum proposed by MacCullagh in 1839 \cite{maccullagh1839,darrigol2010}---treated here as the low-energy, purely electromagnetic approximation to the deformable spacetime metric of General Relativity. 

The primary advantage of adopting MacCullagh's action is its remarkable mathematical brevity and analytical efficiency.
Rather than treating electric and magnetic fields as separate axiomatic entities, the source-free Maxwell equations
emerge directly and kinematically from a single-parameter action governing the continuum's resistance to twist.
This framework cuts through layers of phenomenological abstraction, allowing us to determine strictly from geometry what a
localized photon must look like to propagate losslessly, how it exerts an exact macroscopic torque on birefringent matter,
and how an accelerated charge sheds it. 

To model the mechanics of emission, we must define the object doing the emitting.
In this paper, we adopt the model of the electron as a dynamically bound topological
defect subject to a strict temporal phase condition (a $\pi$-twist) \cite{executable_vacuum}.
The full derivation of this electron model, its precise relationship to General Relativity and the equivalence principle,
and the resolution of historical objections such as Fizeau's optical drag, are detailed in a companion paper \cite{executable_vacuum}.
Here, the electron is treated strictly as a mathematical boundary condition whose co-moving strain field provides the source terms for emission.

The one physical assumption beyond MacCullagh's Lagrangian introduced in this paper is that the medium admits \emph{line defects}: thin cores along which its linear description fails. We do not model the interior of a core. We show instead that the field outside depends on a core only through the circulation round it, allowing us to build the photon from a pair of them. The results therefore hold whether the core is a true discontinuity or a smooth tube.

\paragraph{The Executable Research Record.} Every claim in this paper is registered, with a status and where possible an executable check, in a short Python script, \texttt{twin\_spine\_rx.py}, which accompanies this manuscript. Claims appear in the text as tags such as \cl{D4}. The statuses are identity (mathematics only), postulate (a modelling input), identified (a standard or measured result), derived (computed, with the computation asserted), and open (a known gap; a passing check means the gap is real). Code excerpts in the text are read from the script itself, and the ledger (Appendix~\ref{app:ledger}) lists what each claim ultimately rests on. The format descends from literate programming \cite{knuth1984lp}.

%=====================================================================================
\section{MacCullagh's medium}
\label{sec:medium}
%=====================================================================================

MacCullagh sought an energy for light in crystals that would reproduce Fresnel's
laws with purely transverse waves, and found one that depends only on the curl of
the medium's displacement $\C$:
\spine{maccullagh}{}
\begin{mathbox}
\[
  \mathcal L=\tfrac12\mu_0|\dot\C|^2-\tfrac12(\nabla\times\C)\cdot\varepsilon^{-1}(\nabla\times\C),
  \qquad \D=\nabla\times\C,\quad \Hv=\dot\C,\quad \E=\varepsilon^{-1}\D,
\]
\[
  \nabla\cdot\D\equiv0,\qquad \nabla\times\Hv-\dot\D\equiv0,\qquad
  \text{Euler--Lagrange:}\quad \nabla\times\E=-\mu_0\dot\Hv .
\]
\end{mathbox}
Gauss's and Amp\`ere's laws hold identically and the Euler--Lagrange equation is
Faraday's law \cl{D1} \cl{ID1}: MacCullagh's Lagrangian is the source-free Maxwell
theory of a crystal in dual-potential form. In vacuum $\varepsilon=\varepsilon_0$.
Two properties of the vacuum equation are used repeatedly:
\spine{symmetry}{}
it is invariant under time reversal, and it is linear, so a solution scaled by
$\alpha$ is a solution with $\alpha^2$ times the energy \cl{D2}.

%=====================================================================================
\section{Line defects and the twin-line field}
\label{sec:defects}
%=====================================================================================

\paragraph{Localisation requires a core.} For any field moving rigidly at $c$,
\spine{luminal}{}
\begin{mathbox}
\[
  f=f(x,y,kz-\omega t),\ k=\omega/c\quad\Longrightarrow\quad
  \Box f\equiv\nabla^2f-\frac1{c^2}\frac{\partial^2f}{\partial t^2}=\nabla_\perp^2f
\]
\end{mathbox}
the longitudinal and time derivatives cancel identically \cl{I1}. Such a field solves
the vacuum equation exactly where its transverse profile is harmonic. A profile
that is harmonic and bounded across the whole plane is constant (Liouville
\cl{ID2}), so a transversely localised field moving rigidly at $c$ cannot satisfy
the vacuum equation throughout a cross-section: somewhere it needs a core
\cl{D3}. A smooth beam has none, which is why it diffracts.

We therefore take the medium to admit line defects, and build the photon from two
of them \cl{P1}: thin cores, the \emph{spines}, of opposite displacement-current
strength $\pm\lambda(\theta)$, with $\theta=kz-\omega t$, a transverse vector
$\pm\mathbf R(\theta)$ from the axis. The line joining them is the spine line.

\paragraph{An exact field off the spines.} The transverse potential of two
opposite lines, carried along at $c$, gives
\spine{twinline}{}
\begin{mathbox}
\[
  \Phi=\frac{\lambda(\theta)}2\ln\frac{|\mathbf r_\perp+\mathbf R|^2}{|\mathbf r_\perp-\mathbf R|^2},\qquad
  \phi=\Phi,\quad \mathbf A=(0,0,\Phi/c),\quad \E=-\nabla\phi-\dot{\mathbf A},\quad \mathbf B=\nabla\times\mathbf A
\]
Off the spines: $\partial_zA_z+c^{-2}\partial_t\phi=0$, $E_z=0$, $\nabla\cdot\E=0$,
$\nabla\times\E=-\dot{\mathbf B}$, $\nabla\times\mathbf B=c^{-2}\dot\E$.
\end{mathbox}
an exact vacuum Maxwell solution everywhere off the spines, for any path
$\mathbf R(\theta)$, fixed or turning, and any strength profile $\lambda(\theta)$
\cl{D4}. It is the TEM field of a twin-lead.

\paragraph{No charge.} MacCullagh's representation has no charges in it:
$\nabla\cdot\D=0$ identically. Writing the field of one spine in his variables,
\spine{nocharge}{}
\begin{mathbox}
One spine on the axis, $\lambda(\theta)=\cos\theta$, $\Lambda(\theta)=\sin\theta$ ($\Lambda'=\lambda$):
\[
  \C=\frac{\varepsilon_0\Lambda(\theta)}{k}\,\frac{\hat{\boldsymbol\varphi}}r
  \ \Longrightarrow\ \nabla\times\C=\varepsilon_0\E,\ \ \dot\C=\Hv\ \ (r>0),\qquad
  \underbrace{\oint\D\cdot\hat{\mathbf n}\,d\ell}_{\text{out of the side}}+\frac{\partial}{\partial z}\underbrace{\oint\C\cdot d\boldsymbol\ell}_{\text{along the core}}=0
\]
\end{mathbox}
the field agrees with the twin-line field off the core, and the net displacement
flux out of any cylinder round the spine is zero \cl{D5}. The flux that the
familiar description ends on a line charge does not end: it turns and runs along
the core, and because that flux varies in time it is a displacement current
flowing along the spine. A spine is a thin tube of longitudinal displacement
current, and `charge' is how its edge looks in a representation that cannot see
it.

\paragraph{The exterior does not see the core.} Let the spine be a smooth tube of
longitudinal displacement of radius $a$, joining the exterior smoothly, and try two
different profiles:
\spine{core}{two core profiles}
\begin{mathbox}
\[
  \C=A(\theta)\,\frac{h(r)}r\,\hat{\boldsymbol\varphi},\quad h(r)=1\ (r\ge a),\qquad
  \nabla\cdot\D=0,\qquad \oint_{r\ge a}\C\cdot d\boldsymbol\ell=\iint D_z\,dA=2\pi A(\theta)
\]
\end{mathbox}
All fields are finite on the axis and $\nabla\cdot\D=0$ everywhere. By Stokes'
theorem the circulation of $\C$ round any loop enclosing the core equals the
displacement flux carried along it, so the spine is fixed by its train
$2\pi A(\theta)$, and the two profiles give identical exteriors \cl{D6}. The
results below therefore hold whether a spine is a true discontinuity of the medium
or a smooth tube.

\paragraph{Why two.} One line on its own has transverse energy per length that
diverges far out; opposite lines cancel at large distance:
\spine{minimal}{}
\begin{mathbox}
\[
  \int_a^\infty\Big(\frac2r\Big)^2 2\pi r\,dr=\infty,\qquad
  \int_a^\infty\Big(\frac{4R_0}{r^2}\Big)^2 2\pi r\,dr<\infty,\qquad
  \Phi\ \xrightarrow{\ r\gg R_0\ }\ -\frac{4R_0\cos\vartheta}{r}
\]
\end{mathbox}
A field that is localised, finite far out and carries a transverse dipole needs
zero net line strength and a non-zero moment, so at least two spines of opposite
strength. The twin spine is the minimal case \cl{D7}. This is the field-theoretic
form of a familiar result of waveguide theory: a dispersion-free TEM wave needs at
least two separate conductors \cite{pozar2012} \cl{ID3}.

\paragraph{Nothing binds the spines.} The transverse Maxwell stress of the
twin-line field vanishes identically:
\spine{stress}{}
\begin{mathbox}
\[
  T_{xx}=T_{xy}=T_{yy}=0,\qquad T_{zz}=-\varepsilon_0|\E|^2;\qquad
  f=\frac{\lambda^2}{2\pi\varepsilon_0d}-\frac{\mu_0(\beta c\lambda)^2}{2\pi d}
  =\frac{\lambda^2(1-\beta^2)}{2\pi\varepsilon_0d}\ \xrightarrow{\ \beta\to1\ }\ 0
\]
\end{mathbox}
In the familiar language, two opposite line sources moving together at $\beta c$
attract electrically and repel magnetically, and the two cancel exactly at
$\beta=1$ \cl{D8}. The pair is neutrally bound. Its coherence is kinematic: both
spines belong to one luminal field in a linear, non-dispersive vacuum, and nothing
slows one relative to the other.

%=====================================================================================
\section{Lossless propagation: why the speed of light fixes the shape}
\label{sec:lossless}
%=====================================================================================

A disturbance propagates without loss only if the strain it brings to each point
passes in finite time and leaves the medium at rest. Let a profile travel
unchanged at $v=\beta c$:
\spine{steady}{}
\begin{mathbox}
\[
  g=g(x,y,\zeta),\ \ \zeta=z-vt:\qquad \nabla_\perp^2g+(1-\beta^2)\,\partial_\zeta^2g=0
\]
\end{mathbox}
The speed decides the character of the equation \cl{D9}.
\spine{wake}{}
\begin{mathbox}
\[
  \beta<1:\ g=\Big(r^2+\frac{\zeta^2}{1-\beta^2}\Big)^{-1/2};\qquad
  \beta>1:\ g=\big(\zeta^2-(\beta^2-1)r^2\big)^{-1/2}\ \ \text{in}\ |\zeta|\ge r\sqrt{\beta^2-1}
\]
\end{mathbox}
Slower than light the equation is elliptic and the field reaches ahead and behind
with power-law tails; faster than light it is hyperbolic and a Mach cone trails
behind. Either way a wake is left \cl{D10}.
\spine{atc}{}
At exactly $v=c$ the longitudinal term vanishes and the equation separates into
$\nabla_\perp^2g=0$ in each cross-section. The train along the axis is then free
and can be finite, and the price is the transverse shape: harmonic off its cores,
which is the twin-line field \cl{D11}. After the train has passed a point,
\spine{rest}{}
\begin{mathbox}
\[
  \dot\C_{\rm behind}=0,\qquad
  \nabla\times\C_{\rm behind}=\begin{cases}0 & r\ge a\\ A_{\rm end}\,G(r)\,\hat{\mathbf z} & r<a\end{cases},\qquad
  A_{\rm end}-A_{\rm start}\propto\int\lambda\,d\theta
\]
\end{mathbox}
the medium's velocity is zero, the remaining displacement outside the core carries
no strain, and strain is left inside the core only if the train has a net part.
Whole sinusoidal cycles leave the medium exactly as they found it; half a cycle
does not \cl{D12}. Finally the energy flux is $c$ times the energy density, along
the axis, at every point \cl{D13}:
\spine{flux}{}
nothing leaves sideways. These together make the propagation lossless.

%=====================================================================================
\section{Two photons: linear and circular}
\label{sec:two}
%=====================================================================================

\begin{figure}[t]
\centering
\includegraphics[width=\textwidth]{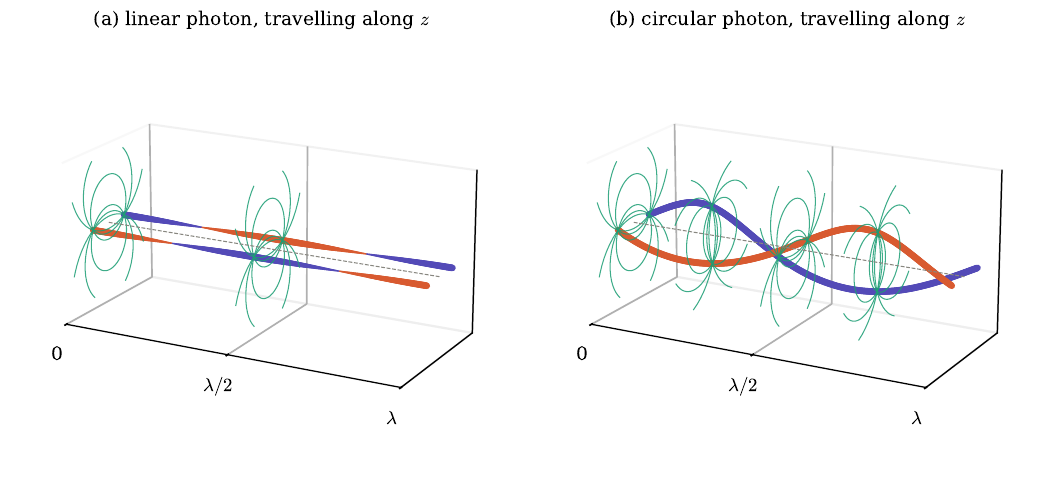}
\caption{The two twin-spine photons over one wavelength (drawn by
\texttt{make\_figures.py}). Purple and orange mark positive and negative
displacement current; line width shows its strength. Green curves in four
cross-sections are the transverse field lines, circles through both spines
\cl{D4}. (a) The linear photon: the spine line is fixed and the strength reverses
through zero each half period \cl{D14}. (b) The circular photon: the strength is
constant and the spine line turns once per period, carrying angular momentum
$E/\omega$ \cl{D20}.}
\label{fig:photon}
\end{figure}

The pair's transverse moment is $\lambda(\theta)\mathbf R(\theta)$, and it is what a
distant absorber sees. Radiation from a bounded source that starts and ends at
rest has no zero-frequency part \cite{jackson1999} \cl{ID4}, so the moment can have
no static part. Compare three structures:
\spine{structure}{}
\begin{mathbox}
\begin{gather*}
  \text{linear:}\quad R_0\cos\theta\binom{\cos\Psi}{\sin\Psi},\qquad
  \text{circular:}\quad R_0\binom{\cos(\theta+\Psi)}{\sin(\theta+\Psi)},\\
  \text{mixed:}\quad R_0\cos\theta\binom{\cos(\theta+\Psi)}{\sin(\theta+\Psi)}
  =\underbrace{\frac{R_0}2\binom{\cos\Psi}{\sin\Psi}}_{\text{static}}+\frac{R_0}2\binom{\cos(2\theta+\Psi)}{\sin(2\theta+\Psi)}
\end{gather*}
\end{mathbox}
The linear pair, spine line fixed at $\Psi$ and strength varying sinusoidally,
oscillates along $\Psi$ alone: the spine line is the polarisation axis. The
circular pair, strength constant and spine line turning once per period, also has
no static part. A pair that both turns and varies in strength rectifies into a
static moment and a doubled frequency, and is excluded \cl{D14}. The two allowed
photons are drawn in Fig.~\ref{fig:photon}. A half twist of the medium has two
geometric readings, and these are they: the dipole reverses by passing through
zero (linear) or by turning through $\pi$ (circular).

\paragraph{The field and the medium.} Between the spines,
\spine{directions}{}
the electric field points along the spine line, while MacCullagh's displacement,
the motion of the medium itself, is perpendicular to it \cl{D15}. This is
MacCullagh's and Neumann's placing of the vibration relative to the plane of
polarisation, as against Fresnel's \cite{darrigol2010}.

\paragraph{A causal scale.} Nothing so far fixes the separation of the spines. We
take it from causality: a spine's displacement current keeps one sign for half a
period, its twisted zone extends $c$ times that, and each spine sits at the edge of
the other's zone \cl{P2}. Then
\spine{scale}{}
\begin{mathbox}
\[
  d=c\,\frac T2=\frac\lambda2,\qquad R_0=\frac\lambda4
\]
\end{mathbox}
the spines are half a wavelength apart, each $\VRnm$\,nm from the axis at
1064\,nm \cl{D16}. No later result depends on $R_0$; it is a prediction
(Section~\ref{sec:open}).

%=====================================================================================
\section{The action, its quantum, and angular momentum}
\label{sec:quantum}
%=====================================================================================

A sinusoidal train of MacCullagh's displacement solves the vacuum equation if and
only if $\omega=ck$ \cl{D17}:
\spine{dispersion}{}
the medium's wave speed fixes the wavelength for a given frequency.
\spine{action}{}
\begin{mathbox}
\[
  \frac1{2\pi}\oint p\,dq=\frac{E}{\omega}\ \ \text{for}\ \ \tfrac12p^2+\tfrac12\omega^2q^2=E;\qquad
  \frac{E_1+E_3}{\omega}-\Big(\frac{E_1}{\omega}+\frac{E_3}{3\omega}\Big)=6\omega\varepsilon^2\ \ \text{for}\ \ \sin\theta+\varepsilon\sin3\theta
\]
\end{mathbox}
A sinusoidal train is one oscillator, whose action variable is $E/\omega$, an
adiabatic invariant; a train with harmonics is several oscillators, and $E/\omega$
is then not its action \cl{D18}. The single-mode sinusoid is the object that has
an action $E/\omega$.

What the action cannot be is fixed by geometry. The medium is linear (D2): scaling
the displacement by $\alpha$ leaves the spine positions, the twist, $R_0$ and the
train unchanged and multiplies the energy by $\alpha^2$. No condition on shape can
therefore fix the energy of a photon \cl{D19}, and the size of the quantum must be
put in. We put it in once: the action $E/\omega$ of a single-mode train comes in
units of $\hbar$ \cl{P3}.

\paragraph{Angular momentum.} Rotations about the axis generate the angular
momentum of MacCullagh's action:
\spine{noether}{}
\begin{mathbox}
\[
  J_z=\int\mu_0\,\dot\C\cdot\delta_\varphi\C\,dV,\qquad
  \delta_\varphi\C=-(x\partial_y-y\partial_x)\C+\hat{\mathbf z}\times\C;
\]
\[
  \text{linear: } j_z=0,\qquad \text{circular: } \frac{j_z\,\omega}{w}=\mp1
\]
\end{mathbox}
For the linear train the displacement and its rate of change stay parallel and
$j_z=0$ pointwise; for the circular train $j_z\omega/w=\mp1$ pointwise. With P3 the
linear photon carries no angular momentum and the circular photon carries $\pm\hbar$
\cl{D20}. One quantum therefore has energy $\hbar\omega$, wavelength
$2\pi c/\omega$, spines $\lambda/2$ apart, and angular momentum $0$ or $\pm\hbar$
according to its geometry \cl{D21}. Angular momentum is quantised by shape once the
action is; energy is not quantised by shape at all.

%=====================================================================================
\section{Birefringence without superposition}
\label{sec:biref}
%=====================================================================================

\paragraph{One path, two stiffnesses.} A birefringent crystal is a medium whose
stiffness differs along two transverse axes. The two Cartesian components of the
one path shared by both spines therefore obey MacCullagh's equation with
different speeds:
\spine{components}{}
\begin{mathbox}
\[
  X=a\cos(n_1k_0z-\omega t),\qquad Y=b\cos(n_2k_0z-\omega t+\phi),\qquad
  \frac{\partial^2q_i}{\partial z^2}-\frac{n_i^2}{c^2}\frac{\partial^2q_i}{\partial t^2}=0
\]
\end{mathbox}
These are components of the path of one object, as the $x$ and $y$ coordinates of
a pendulum bob are components of one motion \cl{D22}. No photon is split, and
nothing is recombined.

\paragraph{The shape of the path.} At a fixed point the spines trace an ellipse once
per period:
\spine{area}{}
\begin{mathbox}
\[
  A=\pi ab\sin\phi,\qquad
  s_1=\frac{a^2-b^2}{a^2+b^2},\ \ s_2=\frac{2ab\cos\phi}{a^2+b^2},\ \ s_3=\frac{2ab\sin\phi}{a^2+b^2},\qquad
  s_1^2+s_2^2+s_3^2=1
\]
\end{mathbox}
Three numbers describe its shape. Its major axis lies at $\tan2\Psi=s_2/s_1$, and its
angular momentum is $J_z\omega/E=s_3=2A/\pi(a^2+b^2)$: zero for the flat path of the
linear photon and $\pm1$ for the circle of the circular photon \cl{D23}. These are
Stokes parameters, here describing the path of a single object.

\paragraph{The grip and the torque.} The crystal's grip on the photon is the stress
$\langle(\D\times\E)_z\rangle$:
\spine{anchor}{}
it equals the rate at which the path gains angular momentum as it opens, so what
the photon gains the crystal gives up \cl{D24}. Across a plate in which the
components progress by $\Delta=k_0\Delta n\,d$ relative to each other,
\spine{retarder}{}
\begin{mathbox}
\[
  \binom{s_2}{s_3}\ \to\ \begin{pmatrix}\cos\Delta&\sin\Delta\\-\sin\Delta&\cos\Delta\end{pmatrix}\binom{s_2}{s_3},
\]
\[
  \frac{\tau}{P/\omega}=s_3^{\rm in}-s_3^{\rm out}
  =\sigma(1-\cos\Delta)+\sqrt{1-\sigma^2}\,\sin\Delta\,\sin2\Psi_0
\]
\end{mathbox}
\begin{equation}
  \tau=\frac{P}{\omega}\Big[\sigma(1-\cos\Delta)+\sqrt{1-\sigma^2}\,\sin\Delta\,\sin2\Psi_0\Big]\qquad\cl{D25}
  \label{eq:master}
\end{equation}
with intercepted power $P$, entry circularity $\sigma=s_3$ and entry orientation
$\Psi_0$. The first term has the sign of $\sigma$ and spins the plate; the second
aligns it. Both are the reaction to a change in the shape of one path.

\begin{figure}[t]
\centering
\includegraphics[width=\textwidth]{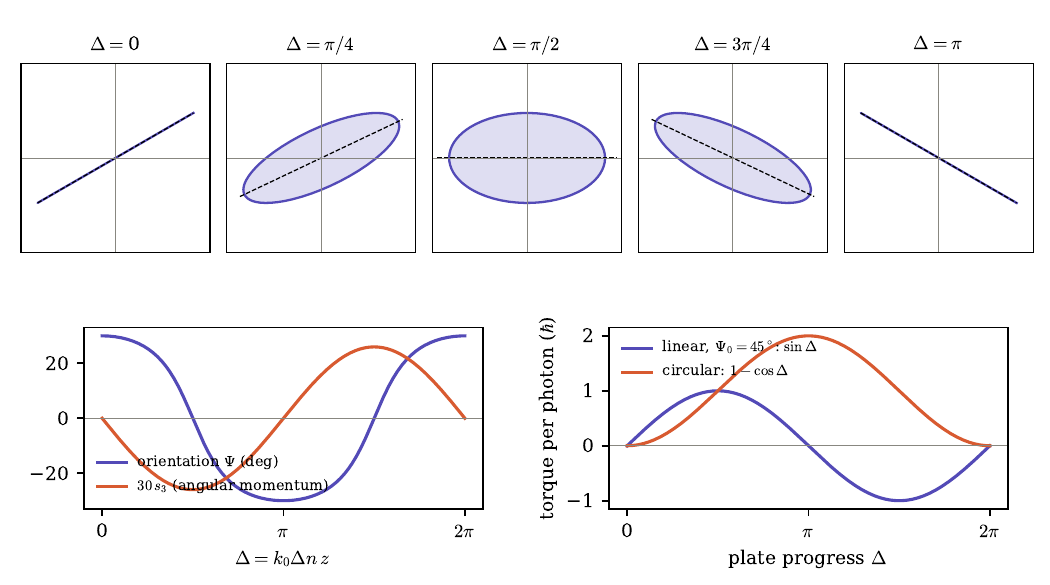}
\caption{Birefringence as a reshaping of one path (drawn by \texttt{make\_figures.py}).
Top: the path traced by the spines of a linear photon entering at $\Psi_0=30^\circ$
to the slow axis, at five depths of progress $\Delta$. The path opens into an
ellipse whose major axis (dashed) swings about the nearer axis \cl{D26}; the shaded
area is the photon's angular momentum \cl{D23}. At $\Delta=\pi$ it is flat again at
$-\Psi_0$ (half-wave plate, \cl{D27}). Bottom left: orientation and angular momentum
with depth. Bottom right: the torque per photon on a plate, Eq.~(\ref{eq:master}),
for a linear photon at $45^\circ$ and a circular photon \cl{D25}.}
\label{fig:biref}
\end{figure}

\paragraph{The linear photon in a plate.} For $\sigma=0$ the path oscillates with
depth (Fig.~\ref{fig:biref}):
\spine{oscillation}{}
\begin{mathbox}
\[
  \tan2\Psi(z)=\tan2\Psi_0\cos\Delta(z),\qquad s_3(z)=-\sin2\Psi_0\sin\Delta(z),\qquad
  \tau=\frac P\omega\sin\Delta\,\sin2\Psi_0
\]
\end{mathbox}
Its orientation swings between $\pm\Psi_0$ about whichever crystal axis is nearer
the entry orientation, while the path opens and closes in step. The pattern
repeats every depth $\lambda_0/\Delta n$; nothing grows without bound \cl{D26}.

\paragraph{Waveplates are path reshapers.}
\spine{waveplates}{}
A half-wave plate, $d=\lambda_0/2\Delta n$, carries the linear photon through half an
oscillation, from $\Psi_0$ to $-\Psi_0$, and leaves it flat. A quarter-wave plate,
$d=\lambda_0/4\Delta n$, with $\Psi_0=45^\circ$, opens the flat path into a circle,
$s_3=-1$: the linear photon leaves as the circular photon of Section~\ref{sec:two},
its spine line now turning, and the plate takes up $\hbar$ \cl{D27}. The two photons
are not two components of one state; they are two shapes of one object, and a
plate converts one into the other.

\paragraph{A recipe for optical tweezers.} Given a particle (birefringence $\Delta n$,
thickness $d$, sphere-equivalent radius $a$) in a fluid of viscosity $\eta$ at
temperature $T$, and a beam of wavelength $\lambda_0$, power $P$ at the particle,
circularity $\sigma$ and orientation $\Psi_0$ to the particle's axes:
(i) $\Delta=2\pi\Delta n\,d/\lambda_0$; (ii) the torque from Eq.~(\ref{eq:master});
(iii) against rotational drag $\gamma_r=8\pi\eta a^3$, circular light spins the particle
at $\tau/\gamma_r$, and linear light aligns it with stiffness
$\kappa_\Psi=2(P/\omega)\sin\Delta$, relaxation time $\gamma_r/\kappa_\Psi$ and thermal
jitter $\sqrt{k_BT/\kappa_\Psi}$.
\spine{example}{}
For a calcite fragment ($n_o=1.642$, $n_e=1.480$ at 1064\,nm \cl{ID5}) with
$d=2.0\,\mu$m, $a=1.5\,\mu$m, in water at 100\,mW: $\Delta=\VDl$\,rad. In linear light
the largest alignment torque is $\Vtaumax$\,N\,m, the stiffness
$\Vkap$\,N\,m\,rad$^{-1}$, the relaxation time $\Vtr$\,ms and the jitter
$\Vpsirms$\,mrad, so the particle is clamped to the spine line. In circular light the
torque is $\Vtauspin$\,N\,m and the particle spins at $\Vfspin$\,Hz \cl{D28}, the
regime of the experiments of Friese \emph{et al.} \cite{friese1998}. A laboratory
comparison must allow for the fraction of power intercepted, shape and wall
corrections to the drag, heating of the water, and the angular spread of a focused
beam.

%=====================================================================================
\section{Emission and absorption}
\label{sec:emission}
%=====================================================================================

We take the electron, in the same spirit as the spines, to be a point defect of the
medium: a core about which the displacement flux $q$ is fixed, carrying with it its
co-moving exterior field \cl{P4} \cite{executable_vacuum}. When the defect is accelerated, the field
outside the sphere $r=ct$ has not yet heard of it and keeps the old configuration;
inside, the new one has formed; the shell between carries the transverse field
$\E_{\rm rad}=(q/4\pi\varepsilon_0c^2r)\,\hat{\mathbf n}\times(\hat{\mathbf n}\times\mathbf a)$
\cite{thomson1904,purcell1965} \cl{ID6}. We ask what this residual looks like
transversely and what can travel away from it.

\begin{figure}[t]
\centering
\includegraphics[width=\textwidth]{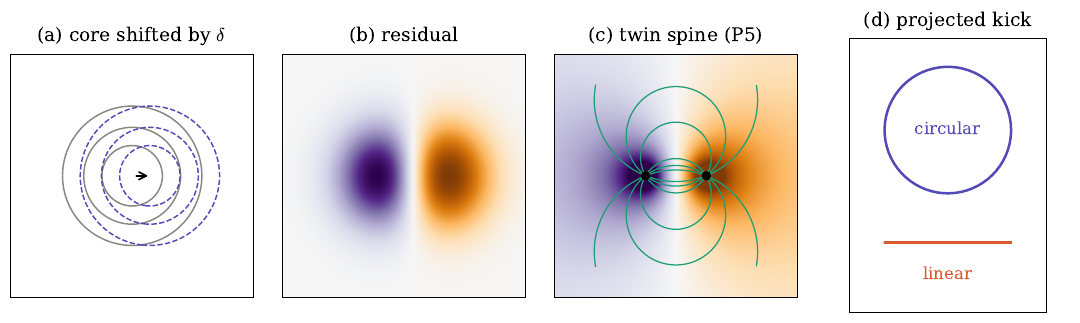}
\caption{Emission (drawn by \texttt{make\_figures.py}). (a) The co-moving field of a
defect before (grey) and after (dashed) a transverse shift of its core. (b) The
residual between them, a smooth transverse dipole aligned with the shift
\cl{D29}. (c) The minimal localised field carrying that dipole at $c$, the twin
spine \cl{D7}; its formation from (b) is postulated (P5), not derived
(\cl{O1}). (d) The projected kick of an electron moving in a circle: circular seen
along its axis, flat seen in its plane, so it sheds circular and linear photons
respectively \cl{D31}.}
\label{fig:emission}
\end{figure}

\paragraph{The residual is a dipole.} Let the co-moving field have a radial
transverse profile $g(r)$, and let the core shift by $\delta$:
\spine{shear}{}
\begin{mathbox}
\[
  G(x-\delta,y)-G(x,y)=-\delta\,\partial_xG+O(\delta^2)=-\delta\,g'(r)\cos\vartheta+O(\delta^2)
\]
\end{mathbox}
The residual between the old and new fields is a pure transverse dipole pattern,
aligned with the shift. It is smooth, and it is not harmonic \cl{D29}
(Fig.~\ref{fig:emission}a,b). By D3 a smooth profile cannot travel rigidly at $c$,
and by D7 the minimal localised field that can, and that carries a transverse
dipole, is the twin spine.

\paragraph{The polarisation axis is the projected kick.} In direction
$\hat{\mathbf n}$,
\spine{projection}{}
\begin{mathbox}
\[
  \hat{\mathbf n}\times(\hat{\mathbf n}\times\mathbf a)=-\mathbf a_\perp,\qquad
  \mathbf a_\perp=\mathbf a-(\hat{\mathbf n}\cdot\mathbf a)\hat{\mathbf n}
\]
\end{mathbox}
the residual is transverse and points along the acceleration projected onto the
plane normal to $\hat{\mathbf n}$. Its moment fixes the spine line of whatever
travels that way \cl{D30}.

\paragraph{The kick history decides the photon.} Follow the projected acceleration
over one period and apply the shape measure of D23:
\spine{kicks}{}
An electron oscillating along a fixed axis gives a flat projected trace in every
direction, and sheds the linear photon. An electron moving in a circle gives a
circular trace along its axis, $J_z\omega/E=\pm1$, and sheds the circular photon
there; in its own plane the trace is flat, and it sheds the linear photon
\cl{D31} (Fig.~\ref{fig:emission}d). This is the observed pattern of $\sigma$ and
$\pi$ light from a single atomic emitter \cite{foot2005}. It is also the
angular-momentum balance: an emitter that loses $\hbar$ about an axis, a
$\Delta m=\pm1$ transition, can shed it only into a circular photon along that axis,
which carries $\pm\hbar$, while a $\Delta m=0$ transition sheds a linear photon,
which carries none \cl{D32}.

\paragraph{Whole cycles.} An emitter that starts and ends at rest has zero net
kick:
\spine{cycles}{}
$\int\mathbf a\,dt=\Delta\mathbf v=0$, so the shed train has no net part. By D12 it is
made of whole cycles and leaves the medium as it found it \cl{D33}. The half twist
that sets the period is half of each cycle; the length of the train is set by how
long the emitter radiates, not by the twist.

\paragraph{Resolution, and what is postulated.} The linear medium spreads the
residual over the dipole pattern $\propto\sin^2$ of the angle to the kick; it does
not gather it into one direction, and it does not form cores. We therefore state
what we assume: an emission event ends in one localised twin-spine pair travelling
in one direction, carrying the projected kick of D30 as its spine line and the
emitter's frequency as its train \cl{P5}. Everything after this postulate is
derived; the mechanism of the step itself is the main open question
(Section~\ref{sec:open}).

\paragraph{Absorption is emission run backwards.} By D2 the time-reversed photon is a
solution of the medium's equation. Arriving at a defect, it reproduces in reverse
the kick history that made it, and delivers its energy and angular momentum to the
defect's motion. Absorbability needs no separate postulate: it follows from the
reversibility of the medium \cl{D34}.

%=====================================================================================
\section{What remains open}
\label{sec:open}
%=====================================================================================

\paragraph{Resolution.} How the smooth residual of an accelerated defect becomes one
localised pair of cores travelling in one direction is not derived (P5). The linear
equation can do neither part; the step requires the defect dynamics of the medium,
which this paper does not supply \cl{O1}.

\paragraph{The core.} The core radius $a$ is a property of the medium. The transverse
energy near a spine grows as $\ln(R_0/a)$, so the energy per length of a given train
depends on it \cl{O2}.

\paragraph{The causal scale as a prediction.} Near-field or single-photon measurements
of the transverse field should find the spines $\lambda/2$ apart. Coherent-beam
torque cannot test this, since Eq.~(\ref{eq:master}) depends only on the shape of the
path \cl{O3}.

\paragraph{The size of the quantum.} Whether $\hbar$ follows from the medium, for
example from the structure of the cores, rather than being put in is open; D19 shows
that in the linear medium it cannot follow from shape alone \cl{O4}.

%=====================================================================================
\section{Conclusion}
%=====================================================================================

If MacCullagh's medium admits line defects, a localised photon follows from his
Lagrangian with little else. A field moving rigidly at $c$ must have cores, and the
minimal arrangement is two of opposite strength, whose field is exact off the cores,
carries no charge and needs no binding. Only at $c$ does the strain a profile brings
pass in finite time and leave the medium at rest. Radiation admits two such photons,
a linear one whose spine line is the polarisation axis and a circular one whose
turning spine line carries $\pm\hbar$, and the action quantum is the one input. In a
crystal the two components of one path travel at different speeds; the reaction to
the path's changing shape is the torque on the crystal, spin and alignment together,
and waveplates convert one photon shape into the other with no superposition. An
accelerated electron leaves a dipole residual whose axis is the projected kick, which
selects the photon shape and balances angular momentum, and absorption is the same
process reversed. The step from that residual to one localised pair is postulated,
and it is where the physics of the defects must next be supplied.

{\sloppy\paragraph{Code availability.} The script \texttt{twin\_spine.py} reproduces every claim in under a
minute (Python with SymPy, NumPy and SciPy \cite{sympy2017,numpy2020,scipy2020}) and
writes its ledger to \texttt{build/ledger.md}; \texttt{make\_figures.py} draws the
figures \cite{matplotlib2007}. Both accompany this paper and are \copyright\ 2026
Imense Ltd, licensed under the GNU Affero General Public License v3 or later; the
licence text is in the file \texttt{LICENSE} distributed with them. Licensing on
other terms is available from Imense Ltd.\par}

\paragraph{Author contributions.} D.A.S.\ conceived the model, developed the theory
and wrote the Executable Research Record. B.H.\ contributed to discussions on the theoretical framework and model development, as well as to the writing and analysis of the results.
Both authors reviewed and approved the manuscript.

\paragraph{Competing interests.} D.A.S.\ is a director of Imense Ltd, which holds the
copyright in this article and its software and is an industrial partner in a grant
application to UK Research and Innovation on this subject. B.H.\ declares no
competing interests.

\paragraph{Acknowledgements.} The authors acknoweldge funding support from Imense Ltd, Cambridge.

\appendix
\section{The claim ledger}
\label{app:ledger}
Generated by \texttt{twin\_spine\_rx.py}; do not edit by hand. \copyright\ 2026 Imense Ltd;
licensed, like the script, under the GNU AGPL v3 or later. `Rests on' lists the
postulates, identifications and open items each claim ultimately depends on. A tick
means the claim's check passes; all \VNclaims\ claims are listed.

\input{build/ledger_table.tex}

\end{document}

%% file: build/values.tex
\newcommand{\VRnm}{266}
\newcommand{\VDl}{1.91}
\newcommand{\Vtaumax}{5.3\times10^{-17}}
\newcommand{\Vkap}{1.1\times10^{-16}}
\newcommand{\Vtr}{0.8}
\newcommand{\Vpsirms}{6.2}
\newcommand{\Vtauspin}{7.5\times10^{-17}}
\newcommand{\Vfspin}{142}
\newcommand{\VNclaims}{50}
\expandafter\def\csname clst@ID1\endcsname{IDENTIFIED}
\expandafter\def\csname clst@D1\endcsname{DERIVED}
\expandafter\def\csname clst@D2\endcsname{DERIVED}
\expandafter\def\csname clst@I1\endcsname{IDENTITY}
\expandafter\def\csname clst@ID2\endcsname{IDENTIFIED}
\expandafter\def\csname clst@D3\endcsname{DERIVED}
\expandafter\def\csname clst@P1\endcsname{POSTULATE}
\expandafter\def\csname clst@D4\endcsname{DERIVED}
\expandafter\def\csname clst@D5\endcsname{DERIVED}
\expandafter\def\csname clst@D6\endcsname{DERIVED}
\expandafter\def\csname clst@ID3\endcsname{IDENTIFIED}
\expandafter\def\csname clst@D7\endcsname{DERIVED}
\expandafter\def\csname clst@D8\endcsname{DERIVED}
\expandafter\def\csname clst@D9\endcsname{DERIVED}
\expandafter\def\csname clst@D10\endcsname{DERIVED}
\expandafter\def\csname clst@D11\endcsname{DERIVED}
\expandafter\def\csname clst@D12\endcsname{DERIVED}
\expandafter\def\csname clst@D13\endcsname{DERIVED}
\expandafter\def\csname clst@ID4\endcsname{IDENTIFIED}
\expandafter\def\csname clst@D14\endcsname{DERIVED}
\expandafter\def\csname clst@D15\endcsname{DERIVED}
\expandafter\def\csname clst@P2\endcsname{POSTULATE}
\expandafter\def\csname clst@D16\endcsname{DERIVED}
\expandafter\def\csname clst@D17\endcsname{DERIVED}
\expandafter\def\csname clst@D18\endcsname{DERIVED}
\expandafter\def\csname clst@D19\endcsname{DERIVED}
\expandafter\def\csname clst@P3\endcsname{POSTULATE}
\expandafter\def\csname clst@D20\endcsname{DERIVED}
\expandafter\def\csname clst@D21\endcsname{DERIVED}
\expandafter\def\csname clst@D22\endcsname{DERIVED}
\expandafter\def\csname clst@D23\endcsname{DERIVED}
\expandafter\def\csname clst@D24\endcsname{DERIVED}
\expandafter\def\csname clst@D25\endcsname{DERIVED}
\expandafter\def\csname clst@D26\endcsname{DERIVED}
\expandafter\def\csname clst@D27\endcsname{DERIVED}
\expandafter\def\csname clst@ID5\endcsname{IDENTIFIED}
\expandafter\def\csname clst@D28\endcsname{DERIVED}
\expandafter\def\csname clst@P4\endcsname{POSTULATE}
\expandafter\def\csname clst@ID6\endcsname{IDENTIFIED}
\expandafter\def\csname clst@D29\endcsname{DERIVED}
\expandafter\def\csname clst@D30\endcsname{DERIVED}
\expandafter\def\csname clst@D31\endcsname{DERIVED}
\expandafter\def\csname clst@D32\endcsname{DERIVED}
\expandafter\def\csname clst@D33\endcsname{DERIVED}
\expandafter\def\csname clst@P5\endcsname{POSTULATE}
\expandafter\def\csname clst@D34\endcsname{DERIVED}
\expandafter\def\csname clst@O1\endcsname{OPEN}
\expandafter\def\csname clst@O2\endcsname{OPEN}
\expandafter\def\csname clst@O3\endcsname{OPEN}
\expandafter\def\csname clst@O4\endcsname{OPEN}

%% file: build/ledger_table.tex
% Copyright (C) 2026 Imense Ltd. Generated by twin_spine.py; licensed under the GNU AGPL v3 or later (SPDX: AGPL-3.0-or-later).
{\small\renewcommand{\arraystretch}{1.15}
\begin{longtable}{@{}l l >{\raggedright\arraybackslash}p{0.50\textwidth} >{\raggedright\arraybackslash}p{0.13\textwidth} c@{}}
\toprule ID & Status & Statement & Rests on & \\ \midrule\endhead
\texttt{ID1} & \st{IDENTIFIED} & Source-free Maxwell equations in a non-magnetic anisotropic crystal, principal permittivities $\varepsilon_x,\varepsilon_y,\varepsilon_z$ (vacuum: all $\varepsilon_0$). & \footnotesize -- & -- \\
\texttt{D1} & \st{DERIVED} & MacCullagh's crystal Lagrangian $\tfrac12\mu_0|\dot{\mathbf C}|^2-\tfrac12(\nabla\times\mathbf C)\cdot\varepsilon^{-1}(\nabla\times\mathbf C)$ is Maxwell's theory in dual-potential form: with $\mathbf D=\nabla\times\mathbf C$, $\mathbf H=\dot{\mathbf C}$, Gauss's and Amp\`ere's laws hold identically and the Euler--Lagrange equation is Faraday's law. & \footnotesize ID1 & \checkmark \\
\texttt{D2} & \st{DERIVED} & The vacuum equation $\mu_0\ddot{\mathbf C}+\varepsilon_0^{-1}\nabla\times\nabla\times\mathbf C=0$ is invariant under time reversal, $\mathbf C(\mathbf r,t)\to\mathbf C(\mathbf r,-t)$, and linear: $\alpha\mathbf C$ is a solution whenever $\mathbf C$ is, with energy density scaled by $\alpha^2$. & \footnotesize ID1 & \checkmark \\
\texttt{I1} & \st{IDENTITY} & For any field moving rigidly at $c$, $f(x,y,kz-\omega t)$ with $k=\omega/c$: $\Box f=\nabla_\perp^2f$. & \footnotesize -- & \checkmark \\
\texttt{ID2} & \st{IDENTIFIED} & Liouville's theorem: a function harmonic and bounded on the whole plane is constant. & \footnotesize -- & -- \\
\texttt{D3} & \st{DERIVED} & A transversely localised field moving rigidly at $c$ cannot satisfy the vacuum equation throughout a cross-section: it needs a core where the linear equation fails. Around a thin line core the transverse potential is the 2D $\ln r$, harmonic off the line. & \footnotesize ID2 & \checkmark \\
\texttt{P1} & \st{POSTULATE} & Line defects: MacCullagh's medium admits thin cores (spines) along which its linear description fails, so that the field is a discontinuous or singular solution there. The photon is two such spines of opposite displacement-current strength $\pm\lambda(\theta)$, $\theta=kz-\omega t$, a transverse vector $\pm\mathbf R(\theta)$ from the axis; the line joining them is the spine line. & \footnotesize ID2 & -- \\
\texttt{D4} & \st{DERIVED} & The twin-line field is an exact vacuum Maxwell solution everywhere off the spines, for any path $\mathbf R(\theta)$ of the spines, fixed or turning, and any strength profile $\lambda(\theta)$: Lorenz gauge, $E_z=0$, $\nabla\cdot\mathbf E=0$, Faraday and Amp\`ere--Maxwell all hold. It is the TEM field of a twin-lead. & \footnotesize ID2, P1 & \checkmark \\
\texttt{D5} & \st{DERIVED} & The spines carry no charge. Written as $\mathbf D=\nabla\times\mathbf C$, $\mathbf H=\dot{\mathbf C}$ the same field agrees with D4 off the core, and the net displacement flux out of any cylinder round a spine is zero: the flux the familiar description ends on a `line charge' turns and runs along the core, and its rate of change is a displacement current along the spine. & \footnotesize ID1, ID2, P1 & \checkmark \\
\texttt{D6} & \st{DERIVED} & The exterior does not see the core. Whether the spine is a true discontinuity or a smooth tube of longitudinal displacement of radius $a$ ($\mathbf C=A(\theta)h(r)\hat{\boldsymbol\varphi}/r$, $h=1$ for $r\ge a$), $\nabla\cdot\mathbf D=0$ and, by Stokes' theorem, the circulation $\oint\mathbf C\cdot d\boldsymbol\ell=2\pi A(\theta)$ equals the displacement flux carried along the core. The exterior field depends on this train alone; two different core profiles give identical exteriors. & \footnotesize ID1, ID2, P1 & \checkmark \\
\texttt{ID3} & \st{IDENTIFIED} & A dispersion-free TEM wave needs at least two separate conductors; a single conductor or hollow guide cannot carry one (standard waveguide theory). & \footnotesize -- & -- \\
\texttt{D7} & \st{DERIVED} & Minimality: a single line has transverse energy per length that diverges at large $r$; zero net line strength removes the divergence. A field that is localised, finite far out and carries a transverse dipole therefore needs at least two spines of opposite strength. The twin line is the minimal case; its far field is exactly the 2D dipole $-4R_0\cos\vartheta/r$. & \footnotesize ID2, ID3, P1 & \checkmark \\
\texttt{D8} & \st{DERIVED} & Nothing binds the spines and nothing needs to: the transverse Maxwell stress of the twin-line field vanishes identically, $T_{xx}=T_{xy}=T_{yy}=0$, leaving only $T_{zz}=-\varepsilon_0|\mathbf E|^2$. Equivalently, opposite line sources moving together at $\beta c$ attract and repel with net force $\lambda^2(1-\beta^2)/2\pi\varepsilon_0d$ per length, zero at $\beta=1$. & \footnotesize ID2, P1 & \checkmark \\
\texttt{D9} & \st{DERIVED} & A profile in steady motion at $v=\beta c$, $g(x,y,\zeta)$ with $\zeta=z-vt$, obeys the vacuum equation iff $\nabla_\perp^2g+(1-\beta^2)\,\partial_\zeta^2g=0$. & \footnotesize ID1 & \checkmark \\
\texttt{D10} & \st{DERIVED} & Slower than light the equation is elliptic and a disturbance's field reaches ahead and behind it with power-law tails, so no point returns to rest in finite time; faster than light it is hyperbolic and a Mach cone $|\zeta|\ge r\sqrt{\beta^2-1}$ trails behind. Either way a wake is left. & \footnotesize ID1 & \checkmark \\
\texttt{D11} & \st{DERIVED} & At $v=c$ the equation separates into $\nabla_\perp^2g=0$ at each $\zeta$: the train along the axis is free and can be finite, so the strain at any point passes in finite time, but each cross-section must be harmonic off its cores, which is the twin-line field (D4, D7). & \footnotesize ID1, ID2, ID3, P1 & \checkmark \\
\texttt{D12} & \st{DERIVED} & Behind the train the medium's velocity $\dot{\mathbf C}$ vanishes. What remains outside the core carries no strain; strain is left inside the core only if the train has a net part, $A_{\rm end}-A_{\rm start}\propto\int\lambda\,d\theta$. Whole sinusoidal cycles leave the medium exactly as they found it; half a cycle does not. & \footnotesize ID1, ID2, ID3, P1 & \checkmark \\
\texttt{D13} & \st{DERIVED} & No energy leaves sideways: $\mathbf S=c\,w\,\hat{\mathbf z}$ at every point, so the energy density moves exactly with the profile. & \footnotesize ID1, ID2, ID3, P1 & \checkmark \\
\texttt{ID4} & \st{IDENTIFIED} & Radiation from a bounded source that starts and ends at rest has no zero-frequency part: $\int\mathbf E_{\rm rad}\,dt\propto\Delta\dot{\mathbf p}=0$. & \footnotesize -- & -- \\
\texttt{D14} & \st{DERIVED} & The transverse moment of the pair is $\lambda(\theta)\mathbf R(\theta)$. Two structures have no static part: the linear pair (spine line fixed at $\Psi$, strength $\propto\cos\theta$), whose field oscillates along $\Psi$ alone, so the spine line is the polarisation axis; and the circular pair (strength constant, spine line turning at $\omega$). A pair that both turns and varies in strength carries a static part $\tfrac12R_0(\cos\Psi,\sin\Psi)$ and a doubled-frequency part, and is excluded by ID4. & \footnotesize ID2, ID4, P1 & \checkmark \\
\texttt{D15} & \st{DERIVED} & Between the spines $\mathbf E$ points along the spine line, while MacCullagh's displacement, the motion of the medium, is perpendicular to it: MacCullagh's and Neumann's placing of the vibration, as against Fresnel's. & \footnotesize ID1, ID2, ID4, P1 & \checkmark \\
\texttt{P2} & \st{POSTULATE} & Causal reach: a spine's displacement current keeps one sign for half a period; its twisted zone extends $c$ times that, and each spine sits at the edge of the other's zone. & \footnotesize ID2, P1 & -- \\
\texttt{D16} & \st{DERIVED} & The spines are half a wavelength apart, $d=\lambda/2$, each $R_0=\lambda/4$ from the axis ($\VRnm$\,nm at 1064\,nm). No later result depends on $R_0$. & \footnotesize ID2, P1, P2 & \checkmark \\
\texttt{D17} & \st{DERIVED} & A sinusoidal train of MacCullagh's displacement along the axis solves the vacuum equation iff $\omega=ck$: $\lambda=2\pi c/\omega$. & \footnotesize ID1 & \checkmark \\
\texttt{D18} & \st{DERIVED} & A sinusoidal train is one oscillator, with action variable $E/\omega$, an adiabatic invariant. A periodic train with harmonics is several oscillators, and $E/\omega$ is then not its total action (for $\sin\theta+\varepsilon\sin3\theta$ they differ by $6\omega\varepsilon^2$). Only the single-mode sinusoid has an action $E/\omega$. & \footnotesize ID1 & \checkmark \\
\texttt{D19} & \st{DERIVED} & No condition on shape can fix the energy: by D2, scaling the displacement by $\alpha$ leaves every geometric feature (spine positions, twist, $R_0$, train) unchanged and scales the energy by $\alpha^2$. A quantum of action must be an input. & \footnotesize ID1 & \checkmark \\
\texttt{P3} & \st{POSTULATE} & The action quantum: the action $E/\omega$ of a single-mode train comes in units of $\hbar$ (Planck's condition in Bohr--Sommerfeld form). This is the one quantum input. & \footnotesize ID1 & -- \\
\texttt{D20} & \st{DERIVED} & Noether's theorem for MacCullagh's action: rotations about the axis generate $J_z=\int\mu_0\dot{\mathbf C}\cdot\delta_\varphi\mathbf C\,dV$. For the linear train $j_z=0$ pointwise; for the circular train $j_z\,\omega/w=\mp1$ pointwise. With P3, the linear photon carries no angular momentum and the circular photon carries $\pm\hbar$. & \footnotesize ID1, ID2, ID4, P1, P3 & \checkmark \\
\texttt{D21} & \st{DERIVED} & One quantum: energy $\hbar\omega$, wavelength $2\pi c/\omega$, spines $\lambda/2$ apart (D16), angular momentum $0$ (linear) or $\pm\hbar$ (circular). & \footnotesize ID1, ID2, ID4, P1, P2, P3 & \checkmark \\
\texttt{D22} & \st{DERIVED} & In a crystal the medium's stiffness differs along its two transverse axes (ID1), so the two Cartesian components of the one shared path of the spines obey MacCullagh's equation with speeds $c/n_1$ and $c/n_2$. These are components of one object's path, not two photons; nothing is superposed. & \footnotesize ID1, ID2, ID4, P1 & \checkmark \\
\texttt{D23} & \st{DERIVED} & At a fixed point the spines trace an ellipse once per period, of signed area $A=\pi ab\sin\phi$. Three numbers describe its shape: $s_1=(a^2-b^2)/(a^2+b^2)$, $s_2=2ab\cos\phi/(a^2+b^2)$, $s_3=2ab\sin\phi/(a^2+b^2)$, with $s_1^2+s_2^2+s_3^2=1$; the major axis lies at $\tan2\Psi=s_2/s_1$ and the path's angular momentum is $J_z\omega/E=s_3=2A/\pi(a^2+b^2)$ (flat path $0$, circle $\pm1$, D20). & \footnotesize ID1, ID2, ID4, P1, P3 & \checkmark \\
\texttt{D24} & \st{DERIVED} & The crystal's grip on the photon is the stress $\langle(\mathbf D\times\mathbf E)_z\rangle$, and it equals the rate at which the path gains angular momentum as it opens: what the photon gains, the crystal gives up. & \footnotesize ID1, ID2, ID4, P1, P3 & \checkmark \\
\texttt{D25} & \st{DERIVED} & A plate in which the path's components progress by $\Delta=k_0\Delta n\,d$ relative to each other turns $(s_2,s_3)$ through $\Delta$ and leaves $s_1$ fixed. The angular momentum the photon loses is the torque on the plate, per photon in units of $\hbar$: $\tau=\sigma(1-\cos\Delta)+\sqrt{1-\sigma^2}\,\sin\Delta\,\sin2\Psi_0$, with $\sigma=s_3$ on entry and $\Psi_0$ the entry orientation. The first term spins the plate, the second aligns it. & \footnotesize ID1, ID2, ID4, P1, P3 & \checkmark \\
\texttt{D26} & \st{DERIVED} & For the linear photon ($\sigma=0$) entering at $\Psi_0$, the path's orientation oscillates with depth as $\tan2\Psi=\tan2\Psi_0\cos\Delta(z)$, swinging between $\pm\Psi_0$ about whichever crystal axis is nearer the entry orientation, while the path opens and closes, $s_3=-\sin2\Psi_0\sin\Delta(z)$. The pattern repeats every $\lambda_0/\Delta n$; nothing grows without bound. The torque reduces to $\tau=\sin\Delta\sin2\Psi_0$. & \footnotesize ID1, ID2, ID4, P1, P3 & \checkmark \\
\texttt{D27} & \st{DERIVED} & Waveplates are path reshapers. A half-wave plate ($d=\lambda_0/2\Delta n$) carries the linear photon through half an oscillation, from $\Psi_0$ to $-\Psi_0$, and leaves it flat. A quarter-wave plate ($d=\lambda_0/4\Delta n$) with $\Psi_0=45^\circ$ opens the flat path into a circle, $s_3=-1$: the linear photon leaves as the circular photon of D14, with its spine line now turning, and the plate takes up $\hbar$. & \footnotesize ID1, ID2, ID4, P1, P3 & \checkmark \\
\texttt{ID5} & \st{IDENTIFIED} & Calcite at 1064\,nm: $n_o=1.642$, $n_e=1.480$; rotational drag of a sphere $8\pi\eta a^3$ in water ($\eta=1.0\times10^{-3}$\,Pa\,s, 295\,K). & \footnotesize -- & -- \\
\texttt{D28} & \st{DERIVED} & Worked example (calcite, $d=2.0\,\mu$m, $a=1.5\,\mu$m, 100\,mW at 1064\,nm): $\Delta=\VDl$\,rad. Linear light: largest alignment torque $\Vtaumax$\,N\,m, stiffness $\Vkap$\,N\,m\,rad$^{-1}$, relaxation $\Vtr$\,ms, thermal jitter $\Vpsirms$\,mrad. Circular light: torque $\Vtauspin$\,N\,m, steady spin $\Vfspin$\,Hz. & \footnotesize ID1, ID2, ID4, ID5, P1, P3 & \checkmark \\
\texttt{P4} & \st{POSTULATE} & The electron is a point defect of the medium: a core about which the displacement flux $q$ is fixed (a topological invariant), carrying with it the standard co-moving exterior field. & \footnotesize ID1 & -- \\
\texttt{ID6} & \st{IDENTIFIED} & The field of an accelerated source: outside the sphere $r=ct$ the old field persists, inside it the new one, and the shell between carries the transverse field $\mathbf E_{\rm rad}=\frac{q}{4\pi\varepsilon_0c^2r}\,\hat{\mathbf n}\times(\hat{\mathbf n}\times\mathbf a)$ (J.\,J.\ Thomson; Purcell). & \footnotesize ID1, P4 & -- \\
\texttt{D29} & \st{DERIVED} & Shear of the bow wave: when the core shifts by $\delta$ transversely, the residual between old and new fields is, to first order, $-\delta\,\partial_xG=-\delta\,g'(r)\cos\vartheta$ for a radial profile $g(r)$: a pure transverse dipole ($m=1$) pattern aligned with the shift. The residual is smooth and not harmonic (for $g=e^{-r^2}$, $\nabla_\perp^2\neq0$). & \footnotesize ID1, ID6, P4 & \checkmark \\
\texttt{D30} & \st{DERIVED} & The residual reaching direction $\hat{\mathbf n}$ is transverse and points along $-\mathbf a_\perp$, the acceleration projected onto the plane normal to $\hat{\mathbf n}$. Its transverse moment fixes the spine line of whatever travels that way: the polarisation axis is the projected kick. & \footnotesize ID1, ID2, ID4, ID6, P1, P4 & \checkmark \\
\texttt{D31} & \st{DERIVED} & The kick history decides the photon. An electron oscillating along a fixed axis gives a flat projected trace in every direction: the linear photon. An electron in circular motion gives a circular trace along its axis ($J_z\omega/E=\pm1$, the circular photon) and a flat trace in its plane (the linear photon). This is the observed pattern of $\sigma$ and $\pi$ light from one emitter. & \footnotesize ID1, ID2, ID4, ID6, P1, P3, P4 & \checkmark \\
\texttt{D32} & \st{DERIVED} & Angular-momentum balance: an emitter that loses $\hbar$ about an axis (a $\Delta m=\pm1$ transition) can shed it only into a circular photon along that axis, which carries $\pm\hbar$ (D20, D31); a $\Delta m=0$ transition sheds a linear photon, which carries none. & \footnotesize ID1, ID2, ID4, ID6, P1, P3, P4 & \checkmark \\
\texttt{D33} & \st{DERIVED} & An emitter that starts and ends at rest has zero net kick, $\int\mathbf a\,dt=\Delta\mathbf v=0$, so the shed train has no net part: by D12 it is made of whole cycles and leaves the medium as it found it. & \footnotesize ID1, ID2, ID3, ID4, ID6, P1, P4 & \checkmark \\
\texttt{P5} & \st{POSTULATE} & Resolution: an emission event ends in one localised twin-spine pair travelling in one direction, carrying the projected kick of D30 as its spine line and the emitter's frequency as its train. & \footnotesize ID1, ID2, ID3, ID4, ID6, P1, P4 & -- \\
\texttt{D34} & \st{DERIVED} & Absorption is emission run backwards. By D2 the time-reversed photon is a solution; arriving at a defect it reproduces, in reverse, the kick history that made it, delivering its energy and angular momentum to the defect's motion. Absorbability follows from reversibility and needs no separate postulate. & \footnotesize ID1, ID2, ID3, ID4, ID6, P1, P4, P5 & \checkmark \\
\texttt{O1} & \st{OPEN} & The resolution P5 is not derived. The linear equation cannot turn the smooth residual of D29 into line singularities, and it spreads the residual over the dipole pattern rather than into one direction; the step needs the defect dynamics of the medium, which this paper does not supply. & \footnotesize ID1, ID2, ID3, ID4, ID6, P1, P4, P5 & \checkmark \\
\texttt{O2} & \st{OPEN} & The core radius $a$ is a property of the medium and is not derived. The transverse energy near a spine grows as $\ln(R_0/a)$, so the energy per length of a given train depends on it. & \footnotesize ID1, ID2, ID3, P1 & \checkmark \\
\texttt{O3} & \st{OPEN} & The causal scale D16 is a prediction: near-field or single-photon measurements of the transverse field should find the spines $\lambda/2$ apart. Coherent-beam torque cannot test it, since D25 depends only on the shape of the path. & \footnotesize ID1, ID2, ID4, P1, P2, P3 & \checkmark \\
\texttt{O4} & \st{OPEN} & Whether $\hbar$ follows from the medium, for example from the structure of the defect cores, rather than being put in (P3); D19 shows it cannot follow from shape alone in the linear medium. & \footnotesize ID1, P3 & \checkmark \\
\bottomrule
\end{longtable}}